# Periodicity in New York State COVID-19 Hospitalizations Leveraged from the Variable Bandpass Periodic Block Bootstrap

Asmaa Ahmad
Aahmad4@albany.edu
Edward Valachovic
evalachovic@albany.edu

Department of Epidemiology and Biostatistics, School of Public Health, University at Albany, State University of New York, One University Place, Rensselaer, NY 12144

Abstract:

The outbreak of the SARS-CoV-2 virus, which led to an unprecedented global pandemic, has underscored the critical importance of understanding seasonal patterns. This knowledge is fundamental for decision-making in healthcare and public health domains. Investigating the presence, intensity, and precise nature of seasonal trends, as well as these temporal patterns, is essential for forecasting future occurrences, planning interventions, and making informed decisions based on the evolution of events over time. This study employs the Variable Bandpass Periodic Block Bootstrap (VBPBB) to separate and analyze different periodic components by frequency in time series data, focusing on annually correlated (PC) principal components. Bootstrapping, a method used to estimate statistical sampling distributions through random sampling with replacement, is particularly useful in this context. Specifically, block bootstrapping, a model-independent resampling method suitable for time series data, is utilized. Its extensions are aimed at preserving the correlation structures inherent in PC processes. The VBPBB applies a bandpass filter to isolate the relevant PC frequency, thereby minimizing contamination from extraneous frequencies and noise. This approach significantly narrows the confidence intervals, enhancing the precision of estimated sampling distributions for the investigated periodic characteristics. Furthermore, we compared the outcomes of block bootstrapping for periodically correlated time series with VBPBB against those from more traditional bootstrapping methods. Our analysis shows VBPBB provides strong evidence of the existence of an annual seasonal PC pattern in hospitalization rates not detectable by other methods, providing timing and confidence intervals for their impact.

## 1. Introduction:

The emergence of Coronavirus Disease 2019 (COVID-19), caused by the Severe Acute Respiratory Syndrome Coronavirus 2 (SARS-CoV-2), marked a significant global health challenge when it was first identified in Wuhan, China, at the close of 2019 (Abebe et al, 2020). This highly infectious respiratory illness primarily spreads through respiratory droplets during close-contact interactions, including coughing, sneezing, or talking, adding to its complexity (Disease of the Week COVID-19, 2021). Further complicating matters, asymptomatic carriers can transmit the virus unknowingly. As our understanding of the virus has evolved, so too have the strategies for managing and containing it, necessitating advances in infection control practices worldwide. From 2019 to 2023, continuous analysis, an enhanced comprehension of trends, and the refinement of predictive models have driven significant advancements in the management of SARS-CoV-2 infections.

This study focuses on investigating the periodic nature of COVID-19 hospitalizations in New York, utilizing the VBPBB method. Its objective is to establish confidence intervals for the seasonal average of hospitalizations, emphasizing the significance of interpreting complex time series data, especially during major health crises like the COVID-19 pandemic. To achieve this, the research employs a novel methodological framework to analyze data on hospitalizations and bed availability from New York hospitals, sourced from the Hospital Electronic Response Data System (HERDS) Hospital Survey: COVID-19 Hospitalizations and Beds. This dataset provides a comprehensive view of hospital resource utilization during the pandemic, shedding light on occupancy trends for COVID-19-diagnosed patients.

Furthermore, this study delves into the concept of bootstrapping, a resampling technique introduced by Efron in 1979, which aids in better understanding datasets and the populations they represent by facilitating the estimation of statistical distributions such as means and variances. However, conventional bootstrapping methods may disrupt inherent correlation structures in time series data, where observations are sequentially ordered and potentially correlated. To address this issue, block bootstrapping methods were developed, aiming to preserve correlations in spatio-temporal data by segmenting the time series into blocks for random sampling. The initial concept and methodology for block bootstrapping in the context of time series data analysis were indeed introduced by Carlstein in 1986. Then, Kunsch (1989) later expanded on this by suggesting the use of overlapping blocks to better address serial correlation.

Specifically, the Moving Block Bootstrap technique, as outlined by Brownstone (2000), involves dividing data into blocks, each comprising a set number of consecutive observations, and resampling these blocks. While this technique generally replicates the autocorrelation structure of the original series in the bootstrap samples, it faces a critical limitation in potentially failing to retain key features, such as periodicity. This limitation arises due to the dependency of observations within the same block in the bootstrap samples, while those in separate blocks remain independent. This research tackles these challenges by evaluating the application of block bootstrapping in the context of PC time series, assessing its effectiveness in preserving critical periodic correlation structures for accurate time series analysis in healthcare settings.

The study of Principal Component (PC) processes, initially advanced by Gladyshev in 1961, has made significant progress and found applications across various fields, such as vibroacoustics, mechanics, signal analysis, hydrology, climatology, and econometrics. Various studies have demonstrated the versatility and power of PC analysis in uncovering underlying patterns and associations within complex datasets. Notable examples include the investigation of ozone levels by Tsakiri and Zurbenko (2010), air quality analysis by Kang et al. (2013), and the study of global temperature trends by Ming and Zurbenko (2012). Additionally, atmospheric studies by Zurbenko and Potrzeba (2010), climate research by Zurbenko and Cyr (2011), and diabetes research by Arndorfer and Zurbenko (2017) further illustrate the application of PC analysis in a range of scientific inquiries. Nevertheless, the analysis of PC processes, particularly in estimating asymptotic covariance matrices for parameters of interest, remains a challenge, underscoring the importance of resampling methods like bootstrapping for constructing reliable confidence intervals (Efron, 1979).

In this research, we investigate the seasonal Principal Component (PC) components and their harmonics of COVID-19 hospitalization rates in the United States, employing both the Generalized Seasonal Block Bootstrap (GSBB) method by Dudek (2014) and the Variable Bandpass Periodic Block Bootstrap (VBPBB) methods proposed by Valachovic (2024). Our primary focus centers on comparing the sizes of confidence intervals generated by these methods for the periodic means of the identified PC components (Valachovic, 2020). Consistently, prior research results demonstrate that the VBPBB method excels in producing more precise confidence intervals for the periodic means (Valachovic and Shishova, 2024). This results in smaller confidence intervals for the PC components in New York State's COVID-19 hospitalization rates.

## 2. Methods:

### 2.1 Data Sources and Analysis

COVID-19 hospitalizations in New York are recorded daily from March 26, 2020, to November 6, 2023, sourced from the Health Data NY (HDNY) Hospital Survey: COVID-19 Hospitalizations and Beds (New York State Department of Health 2024). This dataset provides a comprehensive view of hospital resource utilization during the pandemic, shedding light on occupancy trends for COVID-19-diagnosed patients and is seen in Figure 1. These data are changed to hospitalization rates per 1000 after adjusting for population with data available from the United States Census Bureau (2023). Analysis is performed in Rstudio version Version 2023.09.1+494 statistical software and uses the KZFT function in the KZA package as detailed in Close and Zurbenko (2013). In this study, the primary principal component (PC) of interest is identified as the annual or seasonal variation, characterized by fluctuations in U.S. COVID-19 hospitalization rates that adhere to a 365-day cycle. This analysis acknowledges that such patterns may exhibit harmonic patterns. Therefore, to gain a comprehensive understanding of the temporal dynamics, secondary PC components manifesting at fractional intervals of the primary cycle—for example at one-half (approximately every 183 days), one-third (approximately every 122 days), one-fourth (about every 91 days), and one-fifth (around every 73 days)—are also incorporated into our examination. This approach allows for a detailed investigation into the periodicity and harmonics of COVID-19 hospitalization trends, contributing to a nuanced understanding of the pandemic's temporal impact on healthcare systems. This research also explored additional PC components, including possible weekly and monthly cycles, along with their harmonics, within New York State COVID-19 hospitalization data. However, no significant variations corresponding to these periods were found, leading to their exclusion from the study. Both GSBB and VBPBB methods bootstrap 10000 resamples for the periodic mean of these PC components.

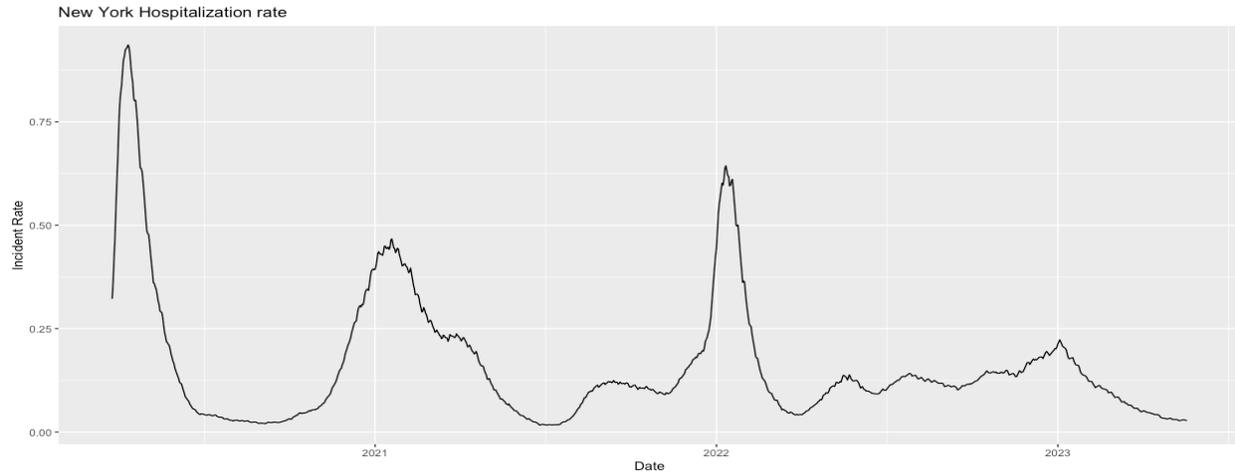

**Figure 1:** NYS COVID-19 Hospitalization rate time series.

## 2.2 The Generalized Seasonal Block Bootstrap method

The Generalized Seasonal Block Bootstrap (GSBB) methodology, introduced by Dudek and colleagues (2014), represents an advanced adaptation of the traditional block bootstrap technique, expressly designed for the analysis of time series data exhibiting periodic correlations. This approach modifies the standard block bootstrap framework to analyze time series more aptly with fixed-length periodicities, facilitating accurate analysis irrespective of the comparative size of the periodicity to the block size or total sample size. The GSBB technique, as further elucidated by Margo (2017), provides a sophisticated methodological tool for precisely capturing seasonal patterns in datasets, enabling the accommodation of periodicities of varying magnitudes relative to the sizes of data blocks and overall samples. Additionally, the GSBB technique permits the segmentation of data into blocks of a specified length, independent of the period's length, and enables the resampling of these blocks in a manner that preserves the inherent periodicity of the dataset (Maiz at el., 2013). Furthermore, the GSBB approach generates a matrix consisting of r=1000 time series, each with a length of n, where n corresponds to the dataset's length. For example, these series are obtained through the resampling of block bootstraps with a period p=365, focusing specifically on the fundamental frequency of the annual principal component while disregarding subsequent components. The 0.975 and 0.025 quantiles from these r resampled time series are determined to establish a 95% confidence interval (CI) for the mean associated with this annual PC component at every time points. This process results in the creation of a GSBB 95% CI band, which profiles the CI range of the periodic mean at each moment throughout the duration of the time series.

## 2.3 Variable Bandpass Periodic Block Bootstrap method

The Variable Bandpass Periodic Block Bootstrap (VBPBB) method, introduced by Valachovic (2024), represents a notable innovation in time series analysis, specifically designed to address the unique challenges of data characterized by periodic components. Like the GSBB and its non-reliance on parametric model assumptions, the VBPBB offers a versatile and sophisticated approach to statistical resampling. VBPBB's foundation is significantly influenced by the periodogram described by Wei (1989), and the concept of iterated moving average bandpass filters, as introduced by Zurbenko in 1986, to effectively identify and isolate specific frequency components. This method targets components at distinct frequencies (or the reciprocal of periods), creating a reconstructed PC time series that retains the essential correlation structure of just one PC component. The subsequent block bootstrapping of this newly formed PC time series, with a carefully chosen block size, is a crucial step in maintaining the correlation structure but resampling only the chosen PC component (Brockwell and Davis, 2002). The role of spectral analysis, particularly the periodogram, is pivotal in uncovering periodic components, thereby enriching the method's capability (Valachovic and Shishova, 2024). VBPBB emerges as a flexible, data-driven solution that more accurately captures the periodicity and correlation structures within time series datasets.

The methodology employed in this study emphasized the uniqueness of the annual principal components, underscoring the significance of the principal component analysis in identifying not only the annual trend but also its harmonic components. Significant components identified by VBPBB contribute substantially to the data's overall variability, rather than being dismissed as random fluctuations. This aspect of the analysis was pivotal, revealing the presence of several influential cycles within the annual period, which are critical for comprehending the temporal dynamics at play in the dataset being examined.

## 2.4 KZFT Filters and Their Roles

The VBPBB method is specifically designed to effectively implement filters on PC time series data. Its primary goal is to segregate and filter the PC time series into distinct spectral density segments, each representing an individual PC component frequency (Valachovic and Shishova, 2024). This strategy is particularly adept when dealing with two or more PC components, using bandpass filters to strategically position their cutoff frequencies between the frequencies associated with the PC components. The use of these filters is crucial for isolating specific frequency components, thereby enhancing the method's ability to investigate the unique properties and characteristics of each PC component.

The VBPBB approach is implemented through the application of the Kolmogorov-Zurbenko Fourier Transform (KZFT), a technique that, as defined in Zurbenko (1986), diverges from traditional Fourier Transforms by integrating elements of moving averages and convolution within the time domain. KZFT is used to decompose a time series into its constituent frequency components, applying a series of weighted moving averages to iteratively smooth the signal and extract frequency content information at various scales. Key functional arguments of the KZFT filters, such as the frequency center ($v$), play a crucial role in this process. The argument $v$ is set to the reciprocal of the period, $1/p$, with $p$ corresponding to period of the PC component of interest, ensuring the filter is symmetrically distributed around this central frequency. Other arguments, like the width of the moving average filter window (defined by argument $m$) and the number of filter iterations (argument $k$), are carefully chosen to define the filter to bandpass only one PC component frequency (Yang and Zurbenko, 2007).

In contrast to the GSBB method, which bootstraps the original time series data without prior frequency separation, VBPBB uniquely focuses on block bootstrapping the frequency-separated PC component time series. This approach involves creating a set of PC time series, generating resamples, and calculating periodic means to capture the average behavior at each cycle of the period $p$. This allows the computation of 95% confidence intervals for the periodic mean from the

0.025 and 0.975 quantiles of the bootstrapped means, thus offering a robust estimate of the PC component.

In this analysis, the VBPBB employs the KZFT functional arguments with the number of iterations (k) set to 1. This is done for consistency across the KZFT filters applied to each Principal Component (PC) and due to the short length of the time series dataset, since higher iterations typically necessitate much longer datasets for effectiveness. For the annual PC components, the window size (m) is set at 731, positioning the edge of the KZFT filter midway between the harmonics, with the filter centered at a frequency ($v$) of 1/365 to target the 365-day or 1-year seasonal component. The same m and k arguments were used for frequencies v=2/365 through v=6/365 to bandpass filter the annual harmonics. Expanding our analytical horizon, the same arguments, m and k, were consistently applied to investigate monthly and weekly components. For the monthly components, centering the KZFT filter at v=1/30 to focus on the 30-day or 1-month seasonal component, and similarly applied to frequencies v=2/30 and v=3/30 for the monthly harmonics. Similarly, for the weekly components, we selected an m value of 731, with the filter centered at v=1/7 to isolate the 7-day or 1-week seasonal component, also applying the same m and k settings for v=2/7 and v=3/7 to filter the weekly harmonics. This uniform application of m and k across different temporal scales including annual, monthly, and weekly, ensures a standardized approach in our time series analysis.

A significant PC component is discerned through the analysis of confidence intervals (CIs) derived from block bootstrapping the frequency-separated PC time series data. By generating a substantial number of resamples specifically, B=10,000 in this instance—periodic means are calculated to reflect the average behavior across each cycle for the period $p$. These means form a distribution upon which 95% confidence intervals are based, using the 0.025 and 0.975 quantiles of the bootstrapped periodic means. A PC component is considered significant when its CI band does not encompass zero throughout the CI band, indicating a deviation from what would be expected by random variation alone. This approach provides a statistically robust method to discern the meaningful components within the time series data.

The square of the correlation coefficient, also known as the coefficient of determination ($R^2$), reveals the proportion of variance in the New York State COVID-19 hospitalization rate time series explained by the principal components.

## 3. Results:

In our comprehensive analysis of COVID-19 hospitalization rates from March 26, 2020, to November 6, 2023, we applied the VBPBB approach, emphasizing the importance of the 365-day cycle. The isolated harmonic components, which stand apart from the primary seasonal trend, provided valuable insights, and the VBPBB method was adept at generating confidence intervals (CIs) for their respective mean values, assuming these means held statistical significance. Of the PC components tested, VBPBB found the 365-day fundamental frequency and the third, fourth, and fifth harmonics significant, while the second and sixth harmonics were tested but found to be insignificant. Meanwhile, GSBB identified only the 365-day fundamental frequency as significant. For the monthly components, both VBPBB and GSBB's 95% CI bands suggest that the monthly component is insignificant, leading to the conclusion that the monthly mean variation is zero. Similarly, for the weekly components, the 95% CI bands from VBPBB and GSBB provide sufficient evidence to conclude that the weekly component is also insignificant.

The GSBB approach is represented by the red region in the Figure 2, showing the 95% CI band for the periodic mean of the 365-day component. This relatively wider red region indicated a significant level of uncertainty or variability in the mean estimate. However, when we applied VBPBB, the results, depicted by the blue region in the figure, showed an improved outcome. Bandpass filtering and bootstrapping the individual PC components led to narrower 95% CIs for the periodic mean, suggesting that VBPBB offers a more precise estimate by isolating each frequency band before bootstrapping.

**Analysis of Hospitalization Trends Using VBPBB with Emphasis on Significant Components**

Annual Component

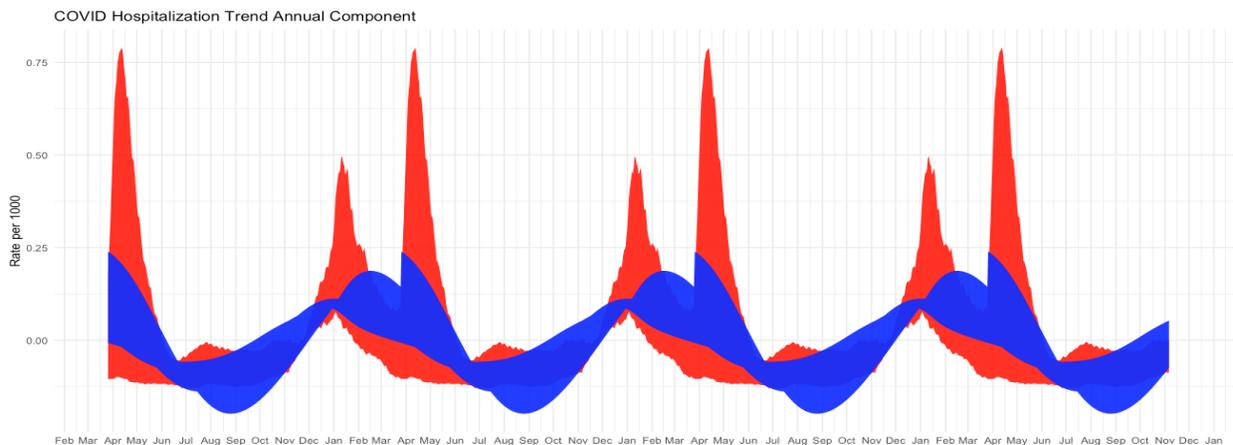

**Figure 2**: 95% CI bands for the NYS COVID-19 Hospitalization rate 365-day, annual, or seasonal mean variation from GSBB in red and VBPBB in blue.

In Figure 2, the graph is showing the 95% confidence intervals bands for the mean hospitalization rates over annual periods. The shaded area in red is the 95% confidence intervals for these annual mean rates for GSBB and in blue for VBPBB. Given the duration of the data (from March 26, 2020, to November 6, 2023), we observed at least three complete seasonal cycles. The peaks correspond to periods of higher hospitalization rates, which could be due to the winter months when respiratory illnesses, including COVID-19, tend to be more prevalent. The troughs, on the other hand, correspond to periods with lower hospitalization rates during the summer months.

GSBB 95% CI for the annual mean ranges from approximately (-0.125, -0.053) additional hospitalization per 1K population to (0.052, 0.491) per 1K population within the annual cycle. VBPBB 95% CI for the annual mean ranges from approximately (-0.138, -0.056) additional hospitalization per 1K population to (0.085, 0.112) per 1K population within the annual cycle. For comparison GSBB, which block bootstraps of the original time series without frequency separation of the PC components, resulted in 95% CIs for the periodic mean that is on average approximately

1.64 times larger than the CIs produced when block bootstrapping the separated PC components using VBPBB. Both the VBPBB and GSBB 95% CI bands offer strong evidence to reject the hypothesis that the seasonal mean variation is zero. According to both approaches, the data significantly supports the existence of a seasonal component in the variation.

Third Annual Harmonic

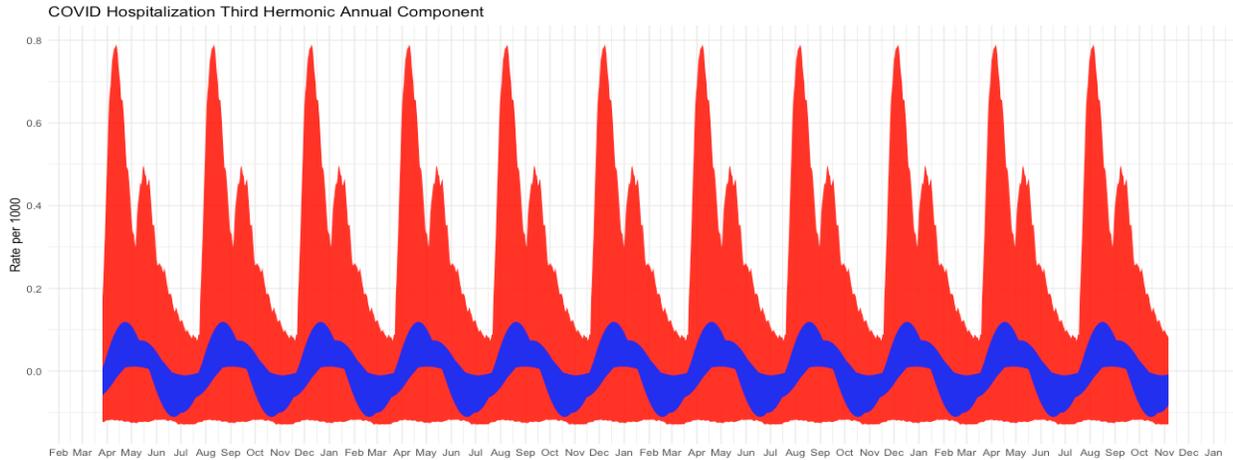

**Figure 3**: 95% CI bands for the NYS COVID-19 hospitalization rate seasonal third harmonic mean variation from GSBB in red and VBPBB in blue.

Given the data spans from March 26, 2020, to November 6, 2023, the third harmonic of the annual component has a periodicity of approximately p=121 days, and we observe a pattern that recurs approximately every four months in Figure 3. The red shaded area shows the variability in the third annual harmonic mean rates using GSBB bootstraps. The GSBB 95% CI for the annual mean ranges from a high of approximately (-0.124, 0.087) additional hospitalization per 1K population to a low of (-0.119, 0.790) per 1K population within the cycle. The blue shaded area reveals the VBPBB 95% CI for the annual mean ranges from approximately (-0.101, -0.010) additional hospitalization per 1K population to (0.009, 0.119) per 1K population within the third annual harmonic cycle. For comparison, block bootstrapping of the PC time series without frequency separation of the PC components, resulted in 95% CIs for the periodic mean that is approximately 5.26 times larger than the CIs produced when block bootstrapping the separated PC components of the PC time series using VBPBB. The VBPBB 95% CI band offers enough evidence to conclude that the third harmonic is significant, rejecting the possibility that the third harmonic of the seasonal mean is zero. However, the GSBB method does not provide enough evidence to support this conclusion.

Fourth Annual Harmonic:

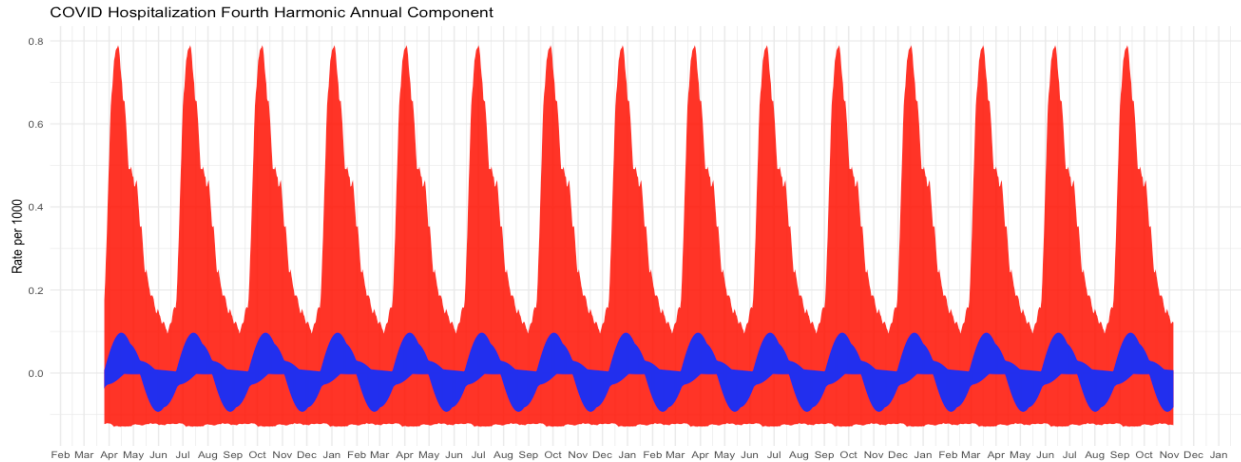

**Figure 4**: 95% CI bands for the NYS COVID-19 hospitalization rate seasonal fourth harmonic mean variation from GSBB in red and VBPBB in blue.

Figure 4 illustrates a pattern that repeats roughly every 89 days, indicating a high-frequency cycle within the annual trend. The red shaded area shows the variability in the fourth annual harmonic mean rates using GSBB bootstraps. GSBB 95% CI for the annual mean ranges from approximately (-0.122, 0.095) additional hospitalization per 1K population to (-0.129, 0.790) per 1K population within the cycle. The blue shaded area shows the VBPBB 95% CI for the annual mean ranges from approximately (0.004, 0.097) additional hospitalization per 1K population to (-0.093, -0.003) per 1K population within the fourth annual harmonic cycle. For comparison, block bootstrapping of the MPC time series without frequency separation of the PC components, resulted in 95% CIs for the periodic mean that is approximately 6.55 times larger than the CIs produced when block bootstrapping the separated PC components of the MPC time series using VBPBB. The VBPBB 95% CI band presents strong evidence in support of the significance of the fourth harmonic, leading to the rejection that the fourth harmonic of the seasonal mean is zero. On the other hand, the GSBB method does not provide this evidence to support that finding.

Fifth Annual Harmonic:

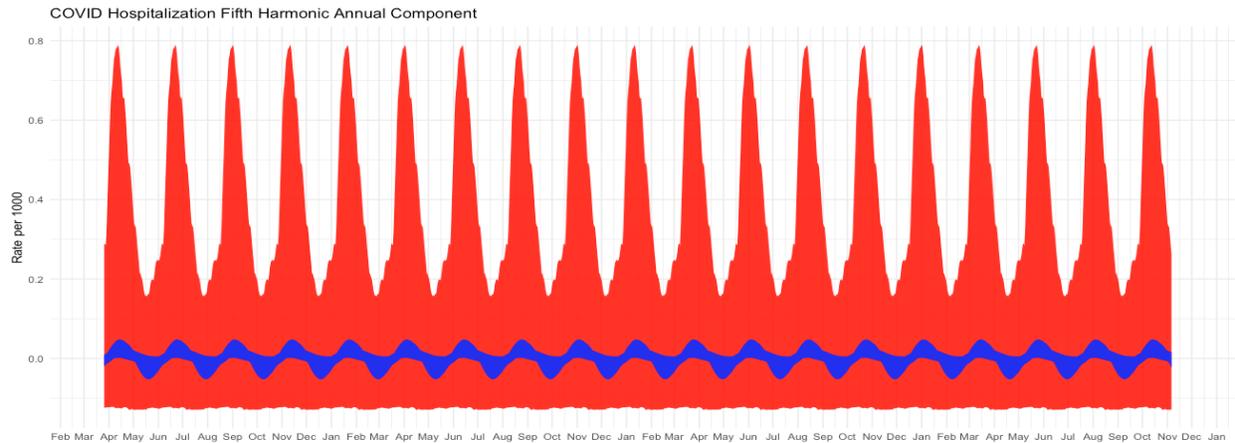

**Figure 5**: 95% CI bands for the NYS COVID-19 hospitalization rate seasonal fifth harmonic mean variation from GSBB in red and VBPBB in blue.

Figure 5 reveals a pattern that repeats approximately every 71 days. The red shaded area shows the variability in the fifth annual harmonic mean rates using GSBB bootstraps. GSBB 95% CI for the annual mean ranges from approximately (-0.127, 0.156) additional hospitalization per 1K population to (-0.124, 0.790) per 1K population within the cycle. The blue shaded area shows the VBPBB 95% CI for the annual mean ranges from approximately (0.0032, 0.0462) additional hospitalization per 1K population to (−0.0541,−0.0004) per 1K population within the fifth annual harmonic cycle. For comparison, block bootstrapping of the MPC time series without frequency separation of the PC components, resulted in 95% CIs for the periodic mean that is approximately 12.36 times larger than the CIs produced when block bootstrapping the separated PC components of the MPC time series using VBPBB. The VBPBB 95% CI band presents strong evidence in support of the significance of the fifth harmonic, leading to the rejection that the fifth harmonic of the seasonal mean is zero. On the other hand, the GSBB method does not support this finding.

**Seasonal and Harmonics**

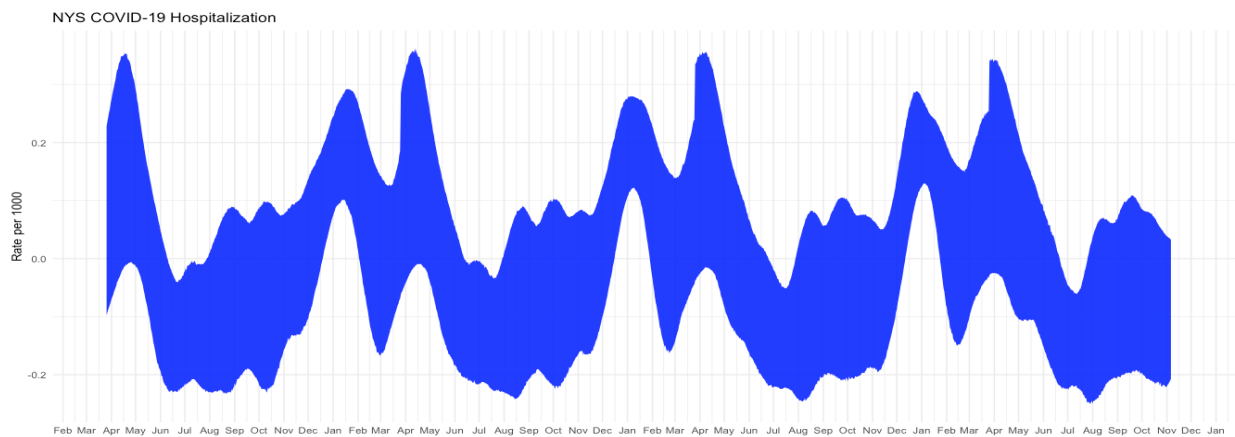

**Figure 6**: The VBPBB 95% CI bands for the NYS COVID-19 hospitalization rate periodic mean variation for seasonality and all significant harmonics are represented in blue.

Figure 6 presents the 95% confidence intervals (CIs) for the mean periodic variation in hospitalization rates due to COVID-19 in New York State, highlighting seasonality and significant harmonic components in blue. These intervals were constructed by combining the first, third, fourth, and fifth principal components. The Variable Bandpass Periodic Block Bootstrap (VBPBB) methodology confirms the importance of these components, as the CI bands notably do not encompass zero, denoting significant periodic mean variations attributable to these elements. The bootstrap method combining the annual component with seasonal harmonic frequencies yields an informative portrayal of seasonal trends, as depicted in Figure 6. Peaks are observed approximately every 150 days, with prominent increases in March, August, and January, and the most pronounced decreases in June and July. Employing the VBPBB to determine annual CI bands for mean variations influenced by these significant principal components provides a powerful tool for enhancing public health strategies against COVID-19.

The first significant principal component accounts for 55% of the variance ($R^2 = .55$), the third accounts for 14% ($R^2 = .14$), the fourth accounts for 0.8% ($R^2 = .008$), and the fifth accounts for 0.5% ($R^2 = .005$). The bandpass filtration process is effective but does not perfectly separate PC component frequencies, therefore the squared correlation of the aggregated PC components does not equal the sum of the squared correlations. When considering the aggregate effect of these significant principal components, the $R^2$ value is .67, suggesting that collectively, they explain 67% of the variability in COVID-19 hospitalization rates.

## 4. Discussion:

This investigation into COVID-19 hospitalization data in New York State, covering March 26, 2020, to November 6, 2023, utilized the Variable Bandpass Periodic Block Bootstrap (VBPBB) method to provide significant insights. Our analysis revealed that VBPBB yields narrower 95% confidence intervals for PC components, including annual or seasonal harmonics, compared to the Generalized Seasonal Block Bootstrap (GSBB). This precision in estimating hospitalization trends is vital for enhancing public health planning and response strategies during the pandemic.

Our statistical analysis indicated that GSBB's 95% confidence intervals for the fundamental annual mean are, on average, 1.64 times wider than that obtained through VBPBB. This disparity becomes more pronounced with increasing complexity of the harmonics: for the third, fourth, and fifth annual harmonics, GSBB's confidence intervals are 5.26, 6.55, and 12.36 times wider, respectively, than those derived from VBPBB. The narrower confidence intervals associated with VBPBB indicate a more reliable estimation of trends that diverge from standard seasonal patterns, which may be influenced by specific factors related to the pandemic. Such trends are influenced by pandemic-related factors like public health policies and virus transmission dynamics, underscoring the importance of precise data analysis for resource management and hospitalization preparedness.

The analysis revealed that the fundamental principal component alone and the original time series has a coefficient of determination ($R^2$) of approximately 0.55. This suggests that the fundamental annual PC component alone accounted for approximately 55% of the variability in COVID-19 hospitalization rates, indicating a strong seasonal influence on hospitalization trends, with higher rates observed in winter and lower in summer. This finding suggests that annual cycles play a significant role in hospitalization rates, a crucial consideration for healthcare planning.

Despite its advantages, VBPBB faces limitations, notably its reliance on moving averages for filtering the time series. This approach can lead to incomplete analyses at the series' ends where averaging is limited by the length of the dataset, emphasizing the need for large datasets for comprehensive analysis. Unlike GSBB, which effectively does not alter data prior to bootstrapping, VBPBB's performance heavily depends on the selection of arguments for bandpass filtration and PC estimation. These factors highlight the necessity for further research into optimizing VBPBB's arguments to enhance its performance further.

Given the relatively short four-year span of the COVID-19 data time series, the findings suggest that both VBPBB and GSBB are close to their methodological limits to properly bootstrapping seasonal components. Therefore, these results should be considered preliminary, with the expectation that additional data will refine these results. Future research should aim to explore beyond the four significant annual PCs of COVID-19 hospitalization rates identified in this study. A deeper application of VBPBB could uncover more insights, potentially enriching public health strategies and improving predictive capabilities for future pandemics.

In conclusion, while this study has provided valuable insights into the seasonal and cyclical patterns of COVID-19 hospitalizations, it also highlights the effectiveness of the VBPBB method in analyzing complex epidemiological data. The ability to accurately characterize the dynamics of COVID-19 hospitalizations is imperative for informed decision-making in public health. Further exploration and refinement of the VBPBB method are recommended to deepen our epidemiological understanding of COVID-19 and enhance our collective response strategies.

References:

1. Abebe, E. C., Dejenie, T. A., Shiferaw, M. Y., & Malik, T. (2020). The newly emerged COVID-19 disease: a systemic review. Virology journal, 17(1), 96.

2. Arndorfer, S. and Zurbenko, I. (2017). Time Series Analysis on Diabetes Mortality in the United States, 1999- 2015 by Kolmogorov-Zurbenko Filter. Journal of Biometrics & Biostatistics. 08. 10.4172/2155-6180.1000384.

3. Brockwell, P. J., & Davis, R. A. (Eds.). (2002). Introduction to time series and forecasting. New York, NY: Springer New York.

4. Brownstone, D., & Valletta, R. (2000). The bootstrap and multiple imputations. Internet: http://www. economics. uci. edu/~ dbrownst/bootmi. pdf.

5. Carlstein, E., Do, K. A., Hall, P., Hesterberg, T., & Künsch, H. R. (1998). Matched-block bootstrap for dependent data. Bernoulli, 305-328.

6. Close, B., and Zurbenko, I. (2013). Kolmogorov-Zurbenko adaptive filters. (Version 3). Retrieved from http://cran.r-project.org/web/packages/kza/index.html

7. Dudek, A. E., Leśkow, J., Paparoditis, E., & Politis, D. N. (2014). A generalized block bootstrap for seasonal time series. Journal of Time Series Analysis, 35(2), 89-114.

8. Dudek, A. E. (2018). Block bootstrap for periodic characteristics of periodically correlated time series. Journal of Nonparametric Statistics, 30(1), 87-124.

9. Disease of the Week - COVID-19. (2021, December 21). Centers for Disease Control and Prevention.  https://www.cdc.gov/dotw/covid-19/index.html

10. Efron, B. (1979). Bootstrap methods: Another look at the jackknife. The Annals of Statistics, 7(1), 1–26. doi:10.1214/aos/1176344552.

11. Gladyshev, E.G. (1961). Periodically correlated random sequences. Soviet Mathematics, 2, 385–388.

12. Kang, D., Hogrefe, C., Foley, K. L., Napelenok, S. L., Mathur, R., & Rao, S. T. (2013). Application of the Kolmogorov–Zurbenko filter and the decoupled direct 3D method for the dynamic evaluation of a regional air quality model. Atmospheric environment, 80, 58-69. https://doi.org/10.1016/j.atmosenv.2013.04.046

13. Kunsch, H. R. (1989). The Jackknife and the Bootstrap for General Stationary Observations. Annals of Statistics, 17(3), 1217-1241. doi: 10.1214/aos/1176347265.

14. Maiz, S., Bonnardot, F., Dudek, A., & Leśkow, J. (2013). Deterministic/cyclostationary signal separation using bootstrap. IFAC Proceedings Volumes, 46(11), 641-646.

15. Margo, L., & Ekonomi, L. (2017). Some empirical results using bootstrap procedures in periodically correlated time series models. Applied Mathematical Sciences, 11(11), 539-547.

16. Ming, L. and Zurbenko, I. (2012). Restoration of Time-Spatial Scales in Global Temperature Data. American Journal of Climate Change, 1(3), 154-163.

17. New York State Department of Health. (2024, February 24). New York State Statewide COVID-19 Hospitalizations and Beds. Retrieved from https://health.data.ny.gov/Health/New-York-State-Statewide-COVID-19-Hospitalizations/jw46-jpb7/about_data.

18. Tsakiri, K. G., and Zurbenko, I. G. (2010). Determining the main atmospheric factor on ozone concentrations. Meteorol Atmos Phys, 109(3), 129-137.

19. Valachovic, E. (2020). Spatio-Temporal Analysis by Frequency Separation: Approach and Research Design. In JSM Proceedings, Section on Statistics and the Environment. Alexandria, VA: American Statistical Association, 1931-1946.

20. Valachovic, E. (2024). Periodically Correlated Time Series and the Variable Bandpass Periodic Block Bootstrap. arXiv:2402.03491 [stat.ME]. Retrieved from http://arxiv.org/abs/2402.03491.

21. Valachovic, E., & Shishova, E. (2024). Seasonal and Periodic Patterns in US COVID-19 Mortality Using the Variable Bandpass Periodic Block Bootstrap. arXiv:2402.03491 [stat.ME]. Retrieved from http://arxiv.org/abs/2403.06343.

22. Wei, W. (1989). Time Series Analysis: Univariate and Multivariate Methods. Redwood City, California: Addison-Wesley.

23. Yang, W., & Zurbenko, I. (2007). KZFT: Kolmogorov-Zurbenko Fourier Transform and Applications. R-Project.

24. Zurbenko, I. (1986). The Spectral Analysis of Time Series. North-Holland, Amsterdam. 248 pages.

25. Zurbenko, I. and Cyr, D. (2011). Climate fluctuations in time and space. Clim Res, 46(1), 67-76.

26. Zurbenko, I., Potrzeba, A. (2010). Tidal waves in the atmosphere and their effects. Acta Geophys. 58, 356–373. https://doi.org/10.2478/s11600-009-0049-y